\documentclass[12pt]{article}
\usepackage[final]{epsfig}
\usepackage{natbib}
\setcitestyle{numbers,square}
\usepackage{float}
\usepackage[T1]{fontenc}
\usepackage{authblk}
\usepackage{natbib}
\usepackage{latexsym}
\usepackage{comment}
\usepackage{amsthm}
\usepackage{verbatim,color}
\usepackage{graphicx}
\usepackage{amssymb}
\usepackage{amssymb, color}
\usepackage{amsmath}
\usepackage{latexsym}
\usepackage{wrapfig}
\usepackage{subfig}
\usepackage[utf8]{inputenc}
\usepackage{multicol}
\usepackage{url}
\usepackage{siunitx}
\usepackage{hyperref}
\usepackage{doi}
\usepackage{makecell}
\usepackage{enumitem}
\usepackage{bm} 
\renewcommand{\title}[1]{\vbox{\center\bf{\Large{#1}}}\vspace{5mm}}
\renewcommand{\author}[1]{\vbox{\center#1}}
\newcommand{\address}[1]{\vbox{\center\em#1}}

\newcommand*{\wideboxed}[1]{\setlength{\fboxsep}{1ex}%
	\fbox{\m@th$\displaystyle#1$}}
\renewcommand{\bar}{\overline}
\renewcommand{\tilde}{\widetilde}
\renewcommand{\hat}{\widehat}
\renewcommand{\leq}{\leqslant}

\newcommand{\bfb}{{\bf b}}

\newcommand{\bfe}{{\bf e}}

\newcommand{\bfg}{{\bf g}}

\newcommand{\bfn}{{\bf n}}

\newcommand{\bfr}{{\bf r}}

\newcommand{\bft}{{\bf t}}

\newcommand{\bfD}{{\bf D}}

\newcommand{\bfL}{{\bf L}}
\newcommand{\bfI}{{\bf I}}

\newcommand{\bfN}{{\bf N}}
\newcommand{\bfP}{{\bf P}}
\newcommand{\bfQ}{{\bf Q}}
\newcommand{\bfR}{{\bf R}}

\newcommand{\bfT}{{\bf T}}

\newcommand{\beq}{\begin{equation}}
\newcommand{\eeq}{\end{equation}}
\newcommand{\beqs}{\begin{eqnarray}}
\newcommand{\eeqs}{\end{eqnarray}}

\begin{document}

	\begin{center}
%		\hfill \\
%		\hfill \\
%		\vskip 1cm
		
		\title{Geometry and Mechanics of Non-Euclidean Curved-Crease Origami}
		
		\author{Zhixuan Wen$^1$, Tian Yu$^2$ and Fan Feng$^{1,*}$} 
  \address{\small  $^1$Department of Mechanics and Engineering Science, College of Engineering, Peking University, Beijing 100871, China\\
  $^2$Department of Mechanics and Aerospace Engineering,
Southern University of Science and Technology, Shenzhen 518055, China\\
\vspace{5pt}
$^*$ fanfeng@pku.edu.cn
\vspace{10pt}
  }
 \today
	\end{center}

	\begin{abstract}
    Recently there have been extensive theoretical, numerical and experimental works on curved-fold origami. However,
	we notice that a unified and complete geometric framework for describing the geometry and mechanics of curved-fold origami, especially those with nontrivial Gaussian curvature at the crease (non-Euclidean crease), is still absent. Herein we provide a unified geometric framework that describes the shape of a generic curved-fold origami composed of two general strips. 
    The explicit description indicates that four configurations emerge, determined by its spatial crease and configuration branch. Within this geometric framework, we derive the equilibrium equations and study the mechanical response of the curved-crease origami, focusing on  Euler's buckling behavior. Both linear stability analysis and finite element simulation indicate that the overlaid configuration exhibits a lower buckling threshold. To further capture the large deformation behavior efficiently, we develop a bistrip model based on the anisotropic Kirchhoff rod theory, which predicts the main features successfully.  This work bridges the geometry and mechanics of curved-crease origami, offering insights for applications in robotics, actuators, and deployable space structures.

 \end{abstract}

\tableofcontents

\section{Introduction}
Origami is an ancient artform that aims to fold a two-dimensional (2D) sheet of paper into a three-dimensional (3D) complex shape. Owing to its unique geometrical and mechanical properties such as deployability and multistability \cite{waitukaitis2015origami}, origami has been not only an aesthetic art but also a promising candidate for developing engineering applications such as robotics, space structures and metamaterials \cite{boatti2017origami}. The most well-studied origami structures are straight-line origami, which involves folding along patterned straight creases. Miura origami and its generalized version quadrilateral origami \cite{feng2020designs} are typical examples, leading to various theories \cite{izmestiev2017classification,tachi2009generalization} and simulation tools \cite{demaine2017origamizer,liu2017nonlinear,filipov2017bar}, for studying deployability and mechanics. Recently, another type of origami, namely curved-fold origami, has gained increasing attention. Efforts have been made by origamists, computer scientists, mathematicians and physicists to study the geometric evolution, numerical simulation and mechanical response of curved-fold origami \cite{feng2024geometry,yu2025unified}, inspiring engineers to develop corresponding applications such as metamaterials, robotics and actuators \cite{doi:10.1126/scirobotics.aaz6262,doi:10.1126/scirobotics.aat0938, lee2021compliant, PRLmeta,meta3, VR}. Despite tremendous efforts in curved-fold origami, we notice that a unified, complete and simple geometric framework for describing curved-fold origami and studying its mechanics is still absent, which limits further applications and is thus the goal of this paper. In particular, we are interested in the geometry and mechanics of curved-fold origami that have non-trivial Gaussian curvature at the crease (i.e. non-Euclidean crease).

\begin{figure}[h]
	\centering
	\includegraphics[width=\columnwidth]{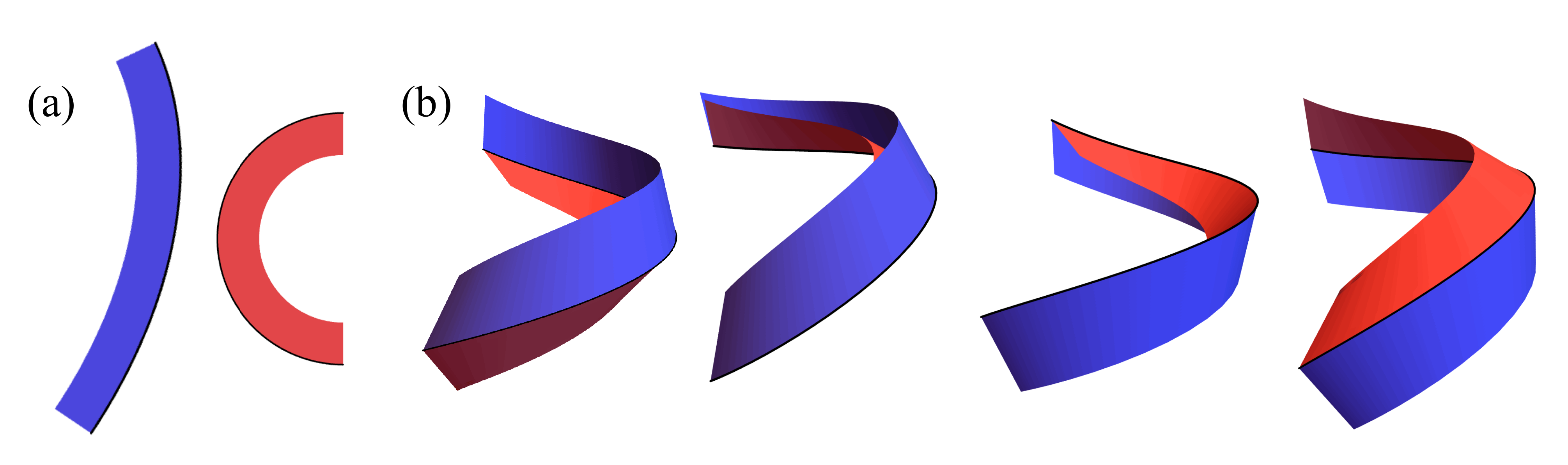}
	\caption{(a) Reference strips. (b) Four configurations indicated by $(\sigma_L, \delta_L, \sigma_R, \delta_R) = (+1,-1,+1,+1), (+1,-1,-1,-1), (-1,+1,+1,+1), (-1, +1, -1, -1)$ with the same spatial curve of the crease. The viewpoints are fixed at the same position.}
	\label{fig:fourconfigurations}
\end{figure}

A curved-fold origami can be made by folding a sheet of paper along a curved crease \cite{liu2024design,liu2025foldable}, or more generically, by stitching two curved strips along their boundaries, as shown in Fig.~\ref{fig:fourconfigurations}. 
Joining two generic strips will impose a continuous boundary condition and also induce non-trivial Gaussian curvature \cite{feng2022interfacial, feng2024geometry} at the crease (please notice that the crease, not the surface \cite{efrati2015non}, is non-Euclidean in this work).
To study its geometric evolution, we draw inspiration from seminal works on thin ribbons and Möbius strips \cite{starostin2007shape,audoly2023analysis,YU2019657, doi:10.1073/pnas.2209048120, starostin2015equilibrium}. 
A thin ribbon favors isometric deformations without stretching, as the bending stiffness is much larger than the stretching stiffness for thin sheets. Therefore, the description of developable surfaces can be adopted if the initial state is flat \cite{do1976differential}. Building upon the shape of a ribbon which is determined by its centerline, previous researchers have obtained important results on the 3D geometry of curved-fold origami in the isometric regime \cite{fuchs1999more}, saying that:
\begin{enumerate}
    \item The shapes of the two flanks are developable surfaces and uniquely determined by the shape of the curved crease in 3D.
    \item The curved crease of the folded state in 3D is more curved than its reference state in 2D to avoid self-penetration. Equivalently, the total curvature of the crease is greater than or equal to the geodesic curvatures.
\end{enumerate}
In this paper, 
we notice that four configurations emerge for a given spatial curved crease (Fig.~\ref{fig:fourconfigurations}(b)) due to the generality of the ``stitching two strips'' approach and the non-trivial Gaussian curvature at the crease. 
We identify all four configurations of a general curved-fold origami under the Frenet frame. The Frenet frame is more natural to use from a mathematical point of view since it is associated only with the curved crease, and we may study the shapes of the two flanks (curved surfaces) under the unified frame. We also find that the four configurations degenerate to two when the reference strips are symmetric.
It should be noted that the Darboux frame, which requires the prior computation of surfaces and is different for the two flanks, is favored by many physicists as it is more convenient to derive the equilibrium equations \cite{dias2012geometric,DIAS201457,WEN2024105829,dias2016wunderlich}, 

Regarding the mechanical response, extensive works have been carried out related to the bending energy of developable surfaces such as ribbons and Möbius strips \cite{starostin2007shape,audoly2023analysis,YU2019657, doi:10.1073/pnas.2209048120,chen2022novel}, which are isometric to a plane (locally) and do not exhibit stretching. 
The full energy density (bending energy per unit area) of a developable surface is given by the Wunderlich functional \cite{Wunderlich1962}, which is proportional to the mean curvature square of the surface. Its degenerate version in the narrow ribbon limit, the Sadowsky functional \cite{sadowsky1930theorie}, is more commonly used to study narrow ribbons, resulting in effective rod theories \cite{YU2019657,doi:10.1073/pnas.2209048120,DIAS201457}. Indeed, a ribbon or a curved-crease origami is a 1D  rod-like object, since the shape is uniquely determined by the centerline or the curved crease subject to certain boundary conditions \cite{fuchs1999more,duncan1982folded,kilian2008curved}. We augment the geometric mechanics theory for Euclidean curved-fold origami \cite{WEN2024105829}, and formulate the equilibrium equations for our non-Euclidean origami, which predicts the shape of the crease accurately. 
A linear stability analysis reveals that the overlaid configuration of the degenerate case has a lower buckling threshold than the symmetric configuration due to the larger configurational space, validated by finite element simulations. To study the mechanics at a low computational cost, we develop a 1D bistrip rod model by simulating the two (narrow) developable surfaces as anisotropic Kirchhoff rods.
The bistrip model can capture the main features of large deformations successfully and efficiently.

This paper is arranged as follows. We first derive the analytical geometric shape of the curved-fold origami under the Frenet frame associated with the curved crease. The folding branches are classified based on their symmetry and Gaussian curvature. Next, equilibrium equations are formulated and Euler's buckling analysis is performed for the degenerate curved-crease origami.
Last, a bistrip rod model is developed based on anisotropic Kirchhoff's rods, which can capture the main features of large deformations efficiently.

\section{Geometry of general non-Euclidean curved-crease origami}\label{sec:geometry}
\begin{figure}[!h]
	\centering
	\includegraphics[width=\columnwidth]{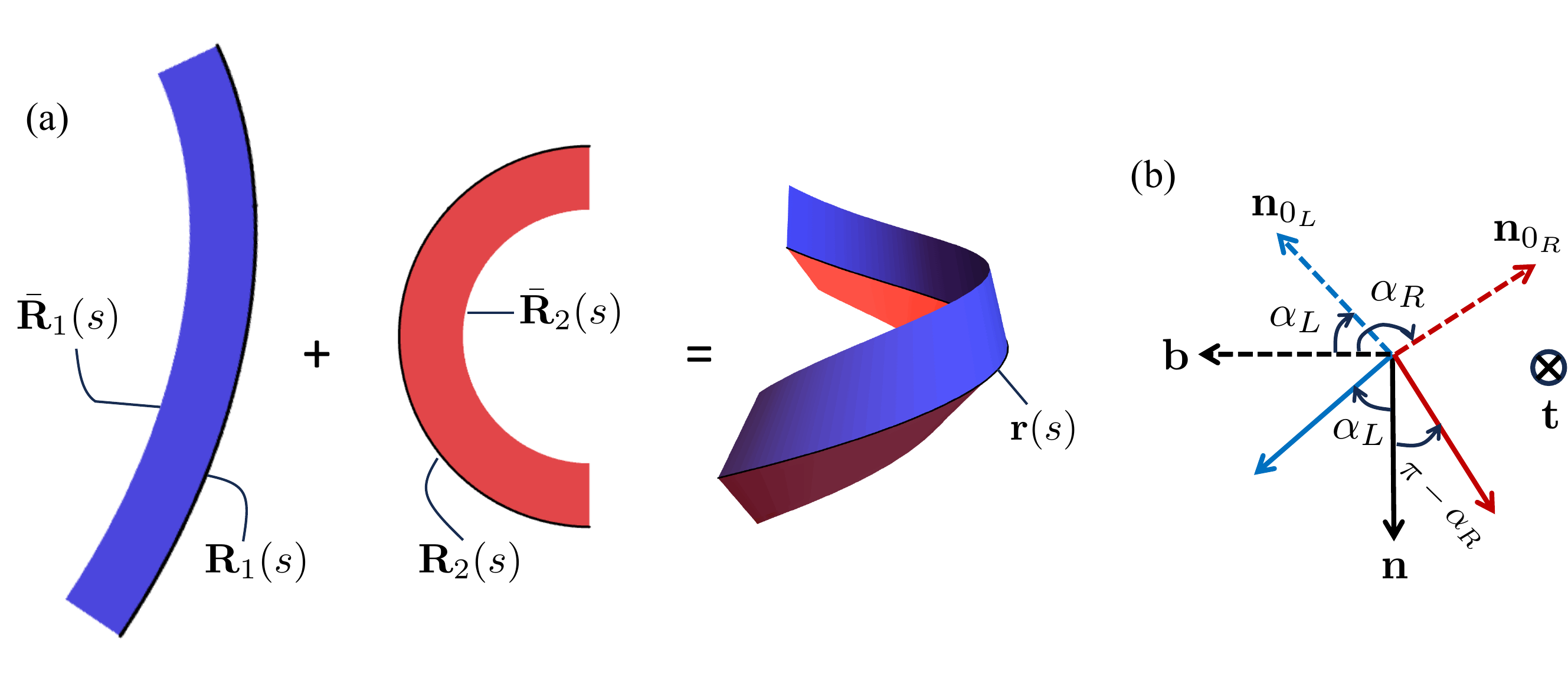}
	\caption{Notations for a general curved-fold origami.}
	\label{fig:notation}
\end{figure}

A general curved-fold origami is constructed by stitching two curved paper strips along their boundaries isometrically (Fig.~\ref{fig:notation}). This approach renders the crease “non-Euclidean”, since the sum of geodesic curvatures is generically nonzero \cite{feng2022interfacial}. Origamists have long appreciated this approach for its versatility and easy fabrication. The fundamental characteristic of this approach is that thin paper is a 2D material that favors bend rather than stretch, making the deformation of each strip isometric to a plane and resulting in each flank as a developable surface. 
Fuchs et al. \cite{fuchs1999more} showed that, after folding a single sheet along a curve, the two flanks are developable surfaces that are determined by the spatial shape of the curved crease.
Fuchs et al. \cite{fuchs1999more} demonstrated that when a single sheet is folded along a curve, the resulting flanks are developable surfaces determined by the spatial shape of the curved crease. Here, we revisit their argument and further derive explicit expressions for the deformed surfaces on both sides of the crease within a unified geometric framework for general curved-fold origami. We will show that, in general, four distinct configurations arise for the same reference strips and spatial crease.

\subsection{Isometric deformations of a single developable surface}
The example in Fig. \ref{fig:notation} is constructed by stitching two reference strips along the arclength parameterized boundaries $\bfR_1(s)$ and $\bfR_2(s)$ with the arclength $s$. 
In the deformed configuration, the two boundaries are mapped to the curved crease $\bfr(s)$ isometrically. 
In the following argument, we will show that given $\bfR_1(s)$, $\bfR_2(s)$ and $\bfr(s)$, the isometric deformations of the deformed (folded) configuration can be analytically derived using the Frenet-Serret frame associated with the crease $\bfr(s)$. 

Specifically, the orthonormal basis of the Frenet-Serret frame is given by
\beq
\bft(s) = \bfr'(s),\quad  \bfn(s) =\bfr''(s)/\kappa(s),\quad \bfb(s) = \bft(s) \times \bfn(s), \label{eq:basis} 
\eeq 
where $\bft(s)$ is the unit tangent, $\bfn(s)$ is the unit normal, $\bfb(s)$ is the binormal, and $\kappa(s)=|\bfr''(s)|$ is the total curvature. The derivatives of the basis vectors with respect to arclength $s$ follow the Frenet-Serret formulae
\beq
\bft'(s)=\kappa(s) \bfn(s), \quad 
\bfn'(s) = -\kappa(s) \bft(s) + \tau(s) \bfb(s),\quad 
\bfb'(s) = -\tau(s) \bfn(s), \label{eq:frenet}
\eeq
with $\kappa(s)$ the curvature and $\tau(s)$ the torsion. 

We first consider the developable surface on the left side of the crease.
The deformation from the flat reference domain to the deformed domain is isometric, thus we have each deformed flank expressed as a ruled surface:
\beq
\bfD(s, u) = \bfr(s) + u \bfg(s), \quad s \in[0,\bar{s}],\quad u \in [0, \bar{u}],
\label{eq:developable_surface}
\eeq
where $\bfg(s)$ is the normalized ruling vector (called generator), $\bar{s}$ is the total arclength of the crease and $\bar{u}$ is the length of the generator at $s$. In addition, the ruled surface must satisfy the
developability condition
\beq
\bfr'(s) \times \bfg(s) \cdot \bfg'(s) = 0
\eeq
to ensure the isometric deformation \cite{o2006elementary}. We then argue that given $\bfr(s)$ and the reference strips, generically there are two solutions $\bfD(s,u)$ that satisfy the developability condition. In the Frenet-Serret frame,  the normalized generator $\bfg(s)$ may be written as 
\beq
\bfg(s) = l_1(s) \bft(s) + l_2(s) \bfn(s) + l_3(s) \bfb(s),\quad l_1(s)^2 + l_2(s)^2 + l_3(s)^2 =1. \label{eq:generator}
\eeq
Substituting $\bfg(s)$ into the developability condition and recalling the derivatives of the basis (\ref{eq:frenet}), we obtain
\beq
\text{\it Developability condition:}\quad
l_2 l_3' - l_3 l_2' - \kappa l_3  l_1 + \tau(l_2^2+ l_3^2) = 0. \label{eq:developability}
\eeq

Meanwhile, the intrinsic property of geodesic curvature yields
\beq
\kappa_g = \kappa  \bfn_0 \cdot \bfb = \kappa \cos\alpha,\label{eq:alpha}
\eeq
where $\kappa_g$ is the geodesic curvature, $\kappa= |\bfr''(s)|$ is the total curvature, 
and $\bfn_0$ is the normal to the tangent plane at the crease.
Then $\alpha = \arccos(\kappa_g/\kappa) \in [0,\pi]$ is the angle between the tangent plane with normal $\bfn_0$ and the osculating plane spanned by $\bft$ and $\bfn$. Due to isometry, the geodesic curvature can also be computed in the reference 2D domain. For example, the left strip has $\kappa_g = \hat{\bfN} \cdot (\bfR'_1(s)\times \bfR''_1(s))$ with $\hat{\bfN}$ the normal to the plane. Substituting the expression of the generator $\bfg$, we have the unit normal to the surface at $\bfr(s)$ as
\beq
\bfn_0(s) = \frac{\bft \times \bfg}{|\bft \times \bfg|}= \frac{l_2 \bfb - l_3 \bfn}{\sqrt{l_2^2 + l_3^2}}. \label{eq:n0}
\eeq

Given the reference strips and deformed crease, $\kappa_g$ and $\kappa$ are prescribed, and thus $\alpha$ is uniquely determined by $\alpha=\arccos(\kappa_g/\kappa)$. However, the normal $\bfn_0$ is not unique. In fact, two solutions of $\bfn_0$ can be obtained as
\beq
\text{\it Intrinsic crease:}\quad
\bfn_0(s) = \cos\alpha \bfb  + \sigma \sin\alpha \bfn, ~\text{with}~\alpha =\arccos(\kappa_g/\kappa),~ \sigma \in \{\pm 1 \}, \label{eq:normal}
\eeq
which results in the two configurations of the left surface mentioned previously.
Taking together the developability condition (\ref{eq:developability}), the intrinsic crease condition (\ref{eq:normal}) and the normal (\ref{eq:n0}), we obtain the explicit expressions of $l_i$ as (omitting the $s$ dependence)
\begin{equation}\label{eq:l1l2l3}
    \begin{aligned}
                l_1 = \delta \frac{ \tau - \sigma \alpha'}{\sqrt{(\tau -\sigma \alpha')^2 + \kappa^2 \sin^2\alpha}}  \\ l_2 = \delta \frac{-\sigma \kappa \sin \alpha \cos\alpha}{\sqrt{(\tau -\sigma \alpha')^2 + \kappa^2 \sin^2\alpha}} \\
l_3 =  \delta \frac{ \kappa \sin^2 \alpha}{\sqrt{(\tau -\sigma \alpha')^2 + \kappa^2 \sin^2\alpha}}  
    \end{aligned}
\end{equation}
where $\delta, \sigma \in \{ \pm 1 \}$ denote the branches of deformation and the term $\kappa \sin \alpha$ is usually referred to as the normal curvature $\kappa_n$. 
We also notice that $\alpha = \arccos(\kappa_g/\kappa)$ gives $\alpha \in [0,\pi]$, resulting in a non-negative $\sin\alpha$, which is
$-\sigma l_3/\sqrt{l_2^2 + l_3^2}$ by computing $\bfn_0 \cdot \bfn$ in Eqs. (\ref{eq:n0}) and (\ref{eq:l1l2l3}).
Substituting the expression of $l_3$ yields $\sigma \delta <0$, and thus we only have two pairs
\beq
(\sigma, \delta) = (+1, -1)~ \text{or}~(\sigma, \delta) = (-1, +1). \label{eq:sign}
\eeq

Taking together (\ref{eq:developable_surface}), (\ref{eq:generator}), (\ref{eq:alpha}),  (\ref{eq:l1l2l3}) and (\ref{eq:sign}), we explicitly obtain the developable surface $\bfL(s,u)=\bfr(s) + u \bfg(s)$, with all the desired quantities determined by $\bfr(s)$ and $\bfR_1(s)$ for the left surface. The two pairs of $(\sigma,\delta)$ correspond to the two configurations. However, the length of the generator $\bar{u}$ remains undetermined. Additionally, the above equation does not exclude the degenerate case in which the generators intersect on the surface. Thus we need the admissible domain to ensure that the deformed configurations are physical.

\subsection{Admissible domain}
\begin{figure}
    \centering
    \includegraphics[width=\linewidth]{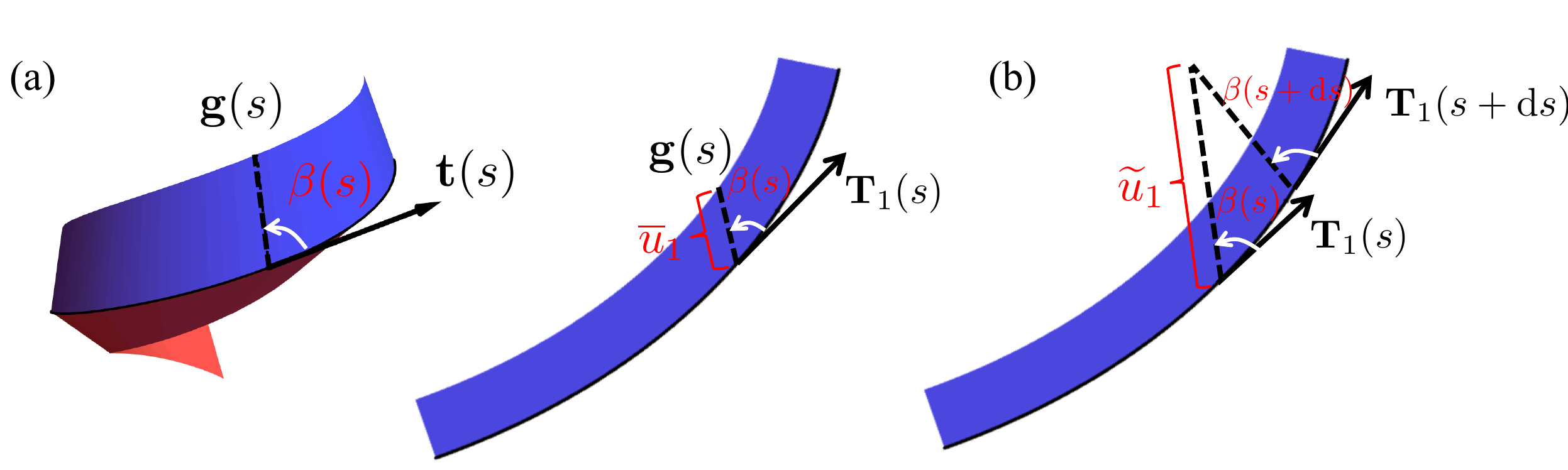}
    \caption{(a) The tangent and generator vectors in the deformed and reference configurations. (b) The intersection of generators.}
    \label{fig:domain}
\end{figure}

In this section, we determine the admissible domain of $u$, i.e., the length of the generator at $\bfr(s)$ and the condition to avoid the intersection of generators. 
First, the angle $\beta(s)$ between the generator $\bfg(s)$ and the tangent of $\bfr(s)$ is preserved before and after deformation, as shown in Fig.~\ref{fig:domain}(a).
Substituting the expression of $\bfg(s)$, the angle $\beta(s)$ can be calculated by 
	\beq
	\beta(s) = \arccos(\bfg(s) \cdot \bft(s)) = \arccos\left(\delta \frac{ \tau - \sigma \alpha'}{\sqrt{(\tau -\sigma \alpha')^2 + \kappa^2 \sin^2\alpha}}\right). \label{eq:beta}
	\eeq
Let the generator $\bfg(s)$ reach the outer boundary $\bar{\bfR}_1(s)$ at $\bar{\bfR}_1(\tilde{s})$. Then the length of generator $\bar{u}_1$ satisfies  
	\beq
	\bfR_1(s) + \bar{u}_1 \bfR(\beta) \bfT_1(s)=\bar{\bfR}_1(\tilde{s}), \label{eq:udomain}
	\eeq
where $\bfR(.) = \cos(.) (\bfe_1\otimes \bfe_1 + \bfe_2 \otimes \bfe_2) + \sin(.)(-\bfe_1\otimes\bfe_2 + \bfe_2 \otimes \bfe_1)$ is a 2D rotation tensor and $\bfT_1(s) = \bfR'_1(s)$ is the tangent. Note that Eq. (\ref{eq:udomain}) is a 2D vector equation with two scalar equations, and thus the two unknowns $\bar{u}_1$ and $\tilde{s}$ are uniquely determined (if the solutions exist).

To avoid the intersection of generators on the surface, the generator $\bfg(s)$ should intersect its neighboring generator $\bfg(s+\mathrm{d} s)$ outside the reference strip in 2D (see Fig.~\ref{fig:domain}(b)). Specifically, 
the intersection of generators satisfies
\beqs
&&\bfR_1(s) + \tilde{u}_1 \bfR(\beta) \bfT_1(s) = \bfR_1(s)(s+\mathrm{d} s) +  (\tilde{u}_1 + \mathrm{d} \tilde{u}_1) \bfR(\beta + \mathrm{d}\beta) \bfT_1(s+\mathrm{d}s) \nonumber \\
\Leftrightarrow  &&\tilde{u}_1' \bfR(\beta) \bfT_1(s) + \tilde{u}_1 \bfR(\beta) \bfT'_1(s) + \tilde{u}_1 \beta' \bfQ\bfR(\beta) \bfT_1(s) + \bfT_1(s) = 0, \label{eq:intersection}
\eeqs
by substituting the derivative of $\bfR(.)$ given by $\bfR'(.) = \bfQ \bfR(.)$ where $\bfQ=- \bfe_1 \otimes \bfe_2 + \bfe_2 \otimes \bfe_1$.
Taking the inner product of (\ref{eq:intersection}) and $\bfR(\beta) \bfT_1^\perp$ yields the solution
\beq
\tilde{u}_1 = -\frac{\bfT_1 \cdot \bfR(\beta)\bfT_1^\perp}{\bfT'_1\cdot\bfT_1^\perp + \beta' \bfQ \bfT_1 \cdot \bfT_1^\perp }  = \frac{\sin\beta}{\kappa_g + \beta'} . \label{eq:threshold}
\eeq

The result of the threshold $\tilde{u}_1$ is consistent with Eq. (9) in Ref. \cite{dias2012geometric}, where the reference boundaries are circles. The sign of $\tilde{u}_1$ depends on $\bfT_1'$, which is not a problem in \cite{dias2012geometric}, since it only considers the circular crease.
In our case, a negative $\tilde{u}_1$  indicates an intersection on the opposite (right) side of the crease, implying that the generators do not intersect on the surface. Therefore, we can redefine the threshold as
\beq
\tilde{u}_1 =
\begin{cases}
	\frac{\sin\beta}{\kappa_g + \beta'}, &\text{if}~\kappa_g + \beta'>0  \\
	+\infty,& \text{if}~\kappa_g + \beta'\leq0
\end{cases},
\eeq
and a sensible length of generator $\bar{u}_1$ must satisfy $\bar{u}_1 \leq \tilde{u}_1$.

\subsection{Geometry of non-Euclidean curved-crease origami}
A non-Euclidean curved-crease origami consists of two developable surfaces.
Following the same procedure, we may derive the expression of the developable surface on the right side of the crease. The difference here is that the normal to the surface is changed to 
\beq
\bfN_R(s, u) = \frac{\bfD_u \times \bfD_s}{|\bfD_u \times \bfD_s|} = \frac{\bfg \times (\bft + u \bfg')}{|\bfg \times (\bft + u \bfg')|}
\eeq
instead of
\beq
\bfN_L(s, u) = \frac{\bfD_s \times \bfD_u}{|\bfD_s \times \bfD_u|} = \frac{(\bft + u \bfg')\times \bfg }{|(\bft + u \bfg')\times \bfg|},
\eeq
and the subsequent calculations need to be changed accordingly. We also realize that the developable surface on the ``right" side of the crease possesses the same form given by (\ref{eq:developable_surface}), (\ref{eq:generator}) and (\ref{eq:l1l2l3}), but with $\sigma \delta = +1$, i.e., $(\sigma, \delta) = (+1, +1)$ or $(\sigma, \delta) = (-1, -1)$. 

We summarize the steps for constructing a curved-crease origami in the deformed configuration. The corresponding Mathematica code is provided in the supplemental material.
\begin{enumerate}
	\item {\bf Reference domain.} The arclength parameterized reference curved interfaces are given by $\bfR_1(s)$ and  $\bfR_2(s)$ with the outer boundary curves $\bar{\bfR}_1(s)$ and $\bar{\bfR}_2(s)$ for the left and right strips, respectively (Fig~\ref{fig:notation}(a)), on the 2D reference domain. The deformed curved crease is given by $\bfr(s)$, with the total curvature
	\beq
	\kappa(s) = |\bfr''(s)| 
	\eeq
	greater than or equal to the absolute value of the geodesic curvature $|\kappa_{gi}(s)|$, $i=L,R$. Here the geodesic curvatures of the crease with respect to both surfaces are given in a unified way as 
	\beq
	\kappa_{g_L}(s)=\hat{\bfN} \cdot (\bfR_1'(s) \times \bfR_1''(s)),~\kappa_{g_R}(s)=\hat{\bfN} \cdot (\bfR_2'(s) \times \bfR_2''(s)),
	\eeq
	where $\hat{\bfN}$ is the unit normal to the reference domain. 
    Notably, in this setting, the Gaussian curvature concentrated at the crease is $\kappa_{g_L} - \kappa_{g_2}$, which is consistent with the setting in Ref. \cite{feng2022interfacial}.
    The surface on the left or the right side of the crease is included in the domain $\{\rho(\hat{\bfN} \times \bfR_1'(s)): ~ \rho>0\}$ or the domain $\{\rho (\bfR_2'(s) \times \hat{\bfN}): ~ \rho>0\}$.
	
	\item {\bf Frenet frame.} Calculate the Frenet-Serret frame associated with $\bfr(s)$ by
	\beq
	\bft(s) = \bfr'(s),~ \bfn(s) =\bfr''(s)/\kappa(s),~ \bfb(s) = \bft(s) \times \bfn(s).
	\eeq 
	The torsion $\tau(s)$ of $\bfr(s)$ is given by
	\beq
	\tau(s)= - \bfb'(s) \cdot \bfn(s). \label{eq:torsion}
	\eeq
	The angles $\alpha(s)$ between the tangent plane of the deformed surface and the osculating plane of $\bfr(s)$, or equivalently, the angles between the normal to the surface and the binormal of the crease, are given by (see Fig.\ref{fig:notation} for notations)
	\beq
	\alpha_L(s) = \arccos[\kappa_{g_L}(s)/\kappa(s)]~\text{and}~ \alpha_R(s) = \arccos[\kappa_{g_R}(s)/\kappa(s)].
	\eeq
	
	\item {\bf Left surface.} The deformed left surface is a developable surface given by
	\beq
	\bfD_L(s, u) = \bfr(s) + u(s) \bfg_L(s),~ s \in[0, \bar{s}],~ u(s) \in [0, u_L(s)] \label{eq:developable_surface_left},
	\eeq
    where $\bar{s}$ is the total arclength of the crease and the generator $\bfg_L$ is given by
	\beq
	\bfg_L(s) = l_{1_L}(s) \bft(s) + l_{2_L}(s) \bfn(s) + l_{3_L}(s) \bfb(s)
	\eeq
	in the Frenet frame, with the coefficients $l_{1_L}, l_{2_L}, l_{3_L}$ (omitting the $s$ dependence)  
	\beqs
	&&l_{1_L} = \delta_L \frac{ \tau - \sigma_L \alpha_L'}{\sqrt{(\tau -\sigma_L \alpha_L')^2 + \kappa^2 \sin^2\alpha_L}},~ l_{2_L} = \delta_L \frac{-\sigma_L \kappa \sin \alpha_L \cos\alpha_L}{\sqrt{(\tau -\sigma_L \alpha_L')^2 + \kappa^2 \sin^2\alpha_L}},  \nonumber \\
	&&l_{3_L} =  \delta_L \frac{ \kappa \sin^2 \alpha_L}{\sqrt{(\tau -\sigma_L \alpha_L')^2 + \kappa^2 \sin^2\alpha_L}}. \label{eq:l1l2l3left}
	\eeqs
	Here $(\sigma_L, \delta_L)=(+1, -1)$ or $(\sigma_L, \delta_L)=(-1, +1)$ determines the two branches of configurations for the left surface. The domain $[0, u_L]$ of $u$ is given by the length of generator $u_L$, which is the solution of 
	\beq
		\bfR_1(s) + \bar{u}_1 \bfR(\beta_L) \bfT_1(s)=\bar{\bfR}_1(\tilde{s}), \label{eq:uleftdomain}
	\eeq
	and also satisfies
	\beq
	u_L \leq \tilde{u}_1=
	\begin{cases}
		\frac{\sin\beta_L}{\kappa_{g_L} + \beta_L'}, &\text{if}~\kappa_{g_L} + \beta_L'>0  \\
		+\infty,& \text{if}~\kappa_{g_L} + \beta_L'\leq0
	\end{cases},
	\eeq
	where $\beta   _L$ is given by Eq.~\ref{eq:beta}.
    
	\item {\bf Right surface.} The deformed right surface is a developable surface given by
	\beq
	\bfD_R(s, u) = \bfr(s) + u(s) \bfg_R(s),~ s\in[0,\bar{s}],~ u(s) \in [0, u_R(s)] \label{eq:developable_surface_right},
	\eeq
	with the generator
	\beq
	\bfg_R(s) = l_{1_R}(s) \bft(s) + l_{2_R}(s) \bfn(s) + l_{3_R}(s) \bfb(s),
	\eeq
	where the coefficients $l_{1_R}, l_{2_R}, l_{3_R}$ (omitting the $s$ dependence) are
	\beqs
	&&l_{1_R} = \delta_R \frac{ \tau - \sigma_R \alpha_R'}{\sqrt{(\tau -\sigma_R \alpha_R')^2 + \kappa^2 \sin^2\alpha_R}},~ l_{2_R} = \delta_R \frac{-\sigma_R \kappa \sin \alpha_R \cos\alpha_R}{\sqrt{(\tau -\sigma_R \alpha_R')^2 + \kappa^2 \sin^2\alpha_R}}, \nonumber \\
	&&l_{3_R} =  \delta_R \frac{ \kappa \sin^2 \alpha_R}{\sqrt{(\tau -\sigma_R \alpha_R')^2 + \kappa^2 \sin^2\alpha_R}}. \label{eq:l1l2l3right}
	\eeqs
	Here $(\sigma_R, \delta_R)=(+1, +1)$ or $(\sigma_R, \delta_R)=(-1, -1)$ determines the two branches of configurations for the right surface. The domain $[0, u_R]$ of $u$ is given the solution $u_R$ of 
	\beq
			\bfR_2(s) + u_R \bfR(-\beta_R) \bfT_2(s)=\bar{\bfR}_2(\tilde{s}), \label{eq:uleftdomain}
	\eeq
	and $u_R$ satisfies
	\beq
	u_R \leq \tilde{u}_2=
	\begin{cases}
		\frac{\sin\beta_R}{-\kappa_{g_R} + \beta_R'}, &\text{if}~-\kappa_{g_R} + \beta_R'>0  \\
		+\infty,& \text{if}~-\kappa_{g_R} + \beta_R'\leq0
	\end{cases}.
	\eeq
    Notice that the minus sign arises due to the difference of `sides' for the left and right surfaces. 
\end{enumerate}
Since each surface has two configurations, the entire curved-crease origami has four configurations in total, as shown in Fig.~\ref{fig:fourconfigurations}(b).

\subsection{A degenerate case with symmetric reference strips}
\begin{figure}[!h]
    \centering
    \includegraphics[width=\linewidth]{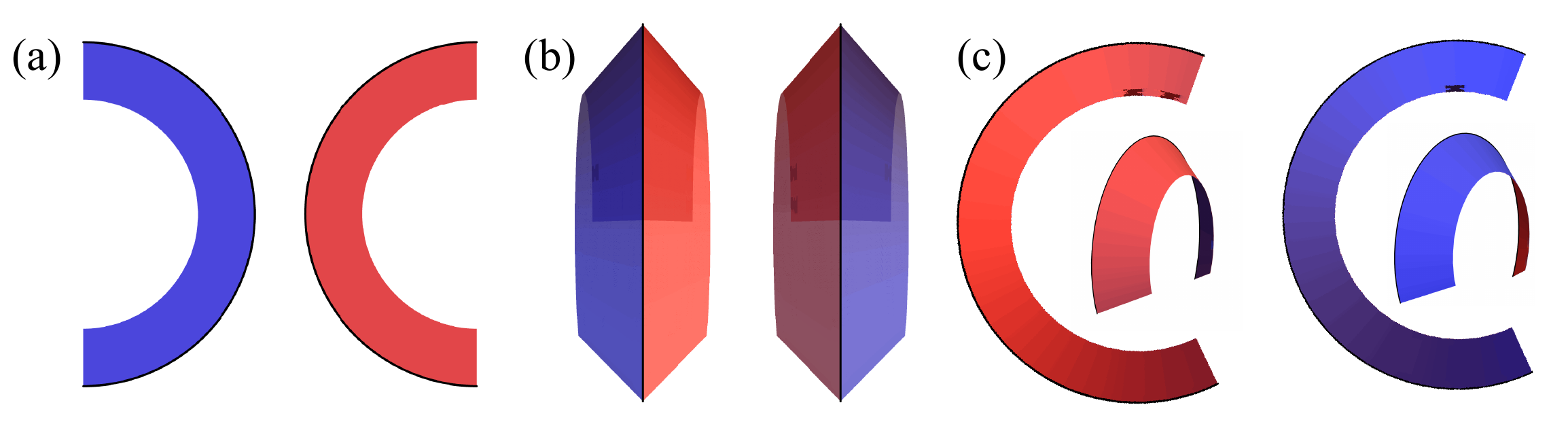}
    \caption{Degenerate curved-crease origami. (a) Reference strips. (b) Symmetric mode: $(\sigma_L, \delta_L, \sigma_R, \delta_R) = (+1,-1,-1,-1), (-1, +1, +1, +1)$. (c) Overlaid mode: $(\sigma_L, \delta_L, \sigma_R, \delta_R) = (+1,-1,+1,+1), (-1,+1,-1,-1)$. Views are from the same position (the insects are side views).}
    \label{fig:symmetric}
\end{figure}

Here we consider a degenerate case composed of two symmetric arcs in the reference configuration. Due to symmetry, only two configurations emerge when the deformed crease remains in 2D as shown in Fig.~\ref{fig:symmetric}, up to rotations.
The symmetric case has geodesic curvatures $\kappa_{g_L} = - \kappa_{g_R}$ in our setting. Assuming the deformed crease remains on a plane, the four solutions in the generic scenario degenerate to two:
\begin{enumerate}
    \item Symmetric mode: $\bfD_R(s,u) - \bfr(s)= (\bfI - 2 \bfb \otimes \bfb) (\bfD_L(s,u) - \bfr(s))$. Notice that $\alpha_L = \alpha_R$ by $\kappa_{g_L} = - \kappa_{g_R}$ and $\tau=0$ for a planar curve. Substituting the expressions of $\bfD_L(s,u)$ and $\bfD_R(s,u)$ with certain $(\sigma_L, \delta_L)$ and $(\sigma_R, \delta_R)$ yields $\bfD_R(s,u) - \bfr(s)= (\bfI - 2 \bfb \otimes \bfb) (\bfD_L(s,u) - \bfr(s))$, corresponding to the symmetric mode shown in Fig.~\ref{fig:symmetric}(b).

    \item Overlaid mode: $\bfD_R(s,u) = \bfD_L(s,u)$. The result follows naturally by submitting the expressions of $\bfD_L(s,u)$ and $\bfD_R(s,u)$ with corresponding $(\sigma_L, \delta_L)$ and $(\sigma_R, \delta_R)$, as the figure shown in Fig.~\ref{fig:symmetric}(c). The two strips are stacked on top of each other.
\end{enumerate}

The curved crease lies on the mirror plane in the symmetric mode, and the two developable surfaces deform symmetrically. In the overlaid mode, the two developable surfaces behave as a single strip. The difference in geometric constraints of these two modes results in different mechanical responses.  As a starting point for the mechanical response, we study the Euler's buckling and large deformation behaviors for the two modes of this degenerate case. More discussions of the general case will follow. 

\section{Mechanics of the curved-crease origami}
In this section, we augment the geometric mechanics theory for Euclidean curved-crease origami in \cite{WEN2024105829} to study the mechanical response (particularly Euler's buckling) of the degenerate non-Euclidean curved-crease origami.
\subsection{Equilibrium equations}
In the isometric deformation regime, the curvature and  torsion of the crease goven the elastic energy of the degenerate curved-crease origami. Following the Sadowsky functional \cite{sadowsky1930theorie}, we have
\begin{equation}
    E=\sum_{\pm}\int_{crease}\frac{1}{2}D\omega(1+t_{\pm}^2)^2\kappa_n^2 \mathrm{d} s=\int_{crease}\varepsilon \mathrm{d} s,
\end{equation}
where $D$ is the bending stiffness (or flexural rigidity),
notations $+,-$ represent the two flanks (left and right), $\omega$ is the panel width and $\varepsilon$ is the equivalent line energy density. $t$ is a function representing the distributions of the generators, defined as
\begin{equation}
    t_{\pm}=\cot \beta_{\pm}=\frac{\kappa_{n\pm}^{\prime}\kappa_{g\pm}}{\kappa^2\kappa_{n\pm}}-\frac{\tau}{\kappa_{n\pm}}.
\end{equation}
%\begin{equation}
    %\varepsilon=D(1+(\frac{\kappa_n^{\prime}\kappa_g}{\kappa^2\kappa_n}-\frac{\tau}{\kappa_n})^2)^2\kappa_n^2.
%\end{equation}
Here $\kappa_{n\pm}$ is the normal curvature of the crease, defined as $\kappa_{n\pm}=\kappa \mathbf{n_{0\pm} \cdot \mathbf{n}}$ ($\bfn_0$ is the normal to the surface and $\bfn$ is the normal vector of the crease). For overlaid mode, $\mathbf{n_{0+}}=-\mathbf{n_{0-}}$, thus $\kappa_{n+}=-\kappa_{n-}$. For symmetric mode, $\mathbf{n_{0+} \cdot \mathbf{n}}=\mathbf{n_{0-} \cdot \mathbf{n}}$, thus $\kappa_{n+}=\kappa_{n-}$. For simplicity, in the following discussion we define $\kappa_n=\kappa_{n+}$.

\begin{figure}[!h]
    \centering
    \includegraphics[width=\linewidth]{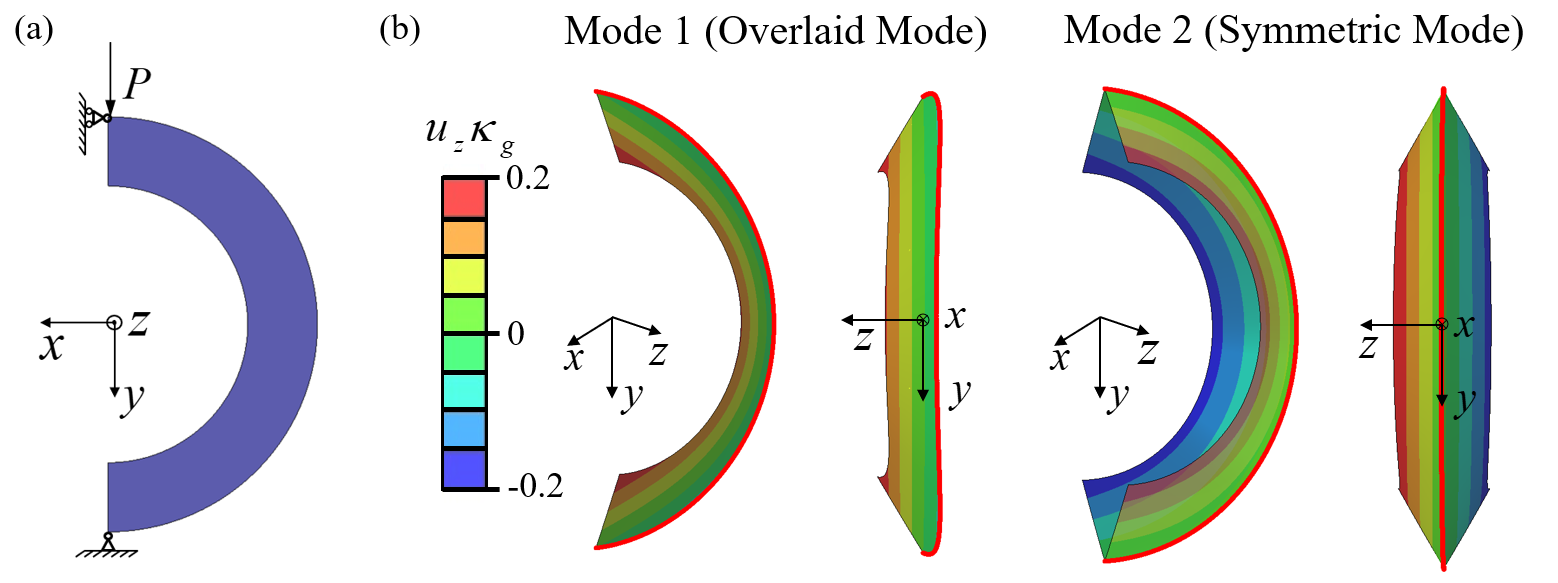}
    \caption{(a) Illustrations for the loading conditions. (b) Deformed configurations of the two buckling modes. The red curves represent creases. The contours represent the displacement in the $z$ direction.}
    \label{fig:FEMshape}
\end{figure}

We consider the Euler's buckling experiment shown in Fig.~\ref{fig:FEMshape}. A degenerate curved-crease origami composed of two arcs is under a compressive force $\bfP$. Passing the threshold, the curved origami will deform either in the overlaid mode (Mode 1) where the two surfaces are on top of each other, or deform in the symmetric mode (Mode 2) where the two surfaces open up symmetrically, assuming the deformation is small.
When compressed, the displacement of both ends of the crease is constrained, the system thus follows the Euler-Lagrange equations given by Eq. (41) in Wen et al. \cite{WEN2024105829}, which are satisfied everywhere along the crease
\begin{equation} \label{eEL}
\begin{aligned}
     f_{gc}&=f\kappa^{\prime}+2\kappa f^{\prime}+(\frac{-f\tau^2-f\kappa^2+f^{\prime\prime}}{\kappa})^{\prime}+g\tau^{\prime}+2\tau g^{\prime}+(\frac{2\tau (\frac{g^{\prime}}{\kappa})^{\prime}+\frac{g^{\prime}}{\kappa} \tau^{\prime}}{\kappa})^{\prime}=0,\\
     g_{gc}&=\frac{g^{\prime}}{\kappa}\tau^2-(g\kappa)^{\prime}-(\frac{g^{\prime}}{\kappa})^{\prime\prime}+f \tau^{\prime}+2\tau f^{\prime}=0,
\end{aligned}
\end{equation}
where $f$ and $g$ are
%\begin{equation}\label{eFG}
%\begin{aligned}
%&f=\partial_{\kappa}\varepsilon-(\partial_{\kappa^{\prime}} \varepsilon)^{\prime}=D\omega(4 \kappa_n^2 t (1+t^2)\partial_{\kappa}t+2\kappa (1+t^2)^2-4 \frac{\kappa_g}{\kappa}(1+3t^2)t^{\prime}), \\
%&g=\partial_{\tau}\varepsilon=-4D\omega t(1+t^2) \kappa_n.
%\end{aligned}
%\end{equation}
\begin{equation}\label{eFG}
\begin{aligned}
&f=\partial_{\kappa}\varepsilon-(\partial_{\kappa^{\prime}} \varepsilon)^{\prime}, \\
&g=\partial_{\tau}\varepsilon.
\end{aligned}
\end{equation}
Following Appendix \ref{sec:appA}, the boundary conditions are also derived by
%\begin{equation}\label{BC}
%\begin{aligned}
%&(\partial_{\kappa^{\prime}}\varepsilon\mathbf{n}+\frac{g}{\kappa}\mathbf{b})\cdot \delta \mathbf{r_c^{\prime\prime}}=0\bigg |_{0}^{s_0}\\
%&(f \mathbf{n}-\frac{g^{\prime}}{\kappa}\mathbf{b})\cdot \delta \mathbf{r_c^{\prime}}\bigg |_{0}^{s_0}=0\\
%&((-\frac{f}{\kappa} \tau^2+\frac{f^{\prime\prime}}{\kappa}+g\tau+\frac{g^{\prime}}{\kappa^2}\tau^{\prime}+2\frac{\tau}{\kappa} (\frac{g^{\prime}}{\kappa})^{\prime})\mathbf{t}
%-(\frac{\tau g^{\prime}}{\kappa}-f^{\prime})\mathbf{n}
%-(f\tau+\frac{g^{\prime}\kappa^{\prime}}{\kappa^2}-\frac{g^{\prime\prime}}{\kappa})\mathbf{b})\cdot\delta \mathbf{r}_c\bigg |_{0}^{s_0}=0
%\end{aligned}
%\end{equation}
\begin{equation}\label{eBC}
\partial_{\kappa^{\prime}}\varepsilon=0,\ \frac{g}{\kappa}=0,\ f=0,\ \frac{g^{\prime}}{\kappa}=0,
\end{equation}
which are satisfied at both ends of the crease.

\subsection{Euler's buckling of the degenerate curved-crease origami}
Under compression, the degenerate curved-crease origami will have different Euler buckling modes at different critical loads, which can be observed in finite element analysis. A degenerate curved-fold origami with $ \omega \kappa_g=1/3$ is modeled, and the configurations of the first and second buckling modes simulated by FEM are shown in Fig. \ref{fig:FEMshape}(b). 

To analyze the buckling modes theoretically, the configurations of the system are solved from Eqs. \eqref{eEL}-\eqref{eBC} under small deformation, i.e., the displacement of the loading point $\delta y \ll \kappa_g^{-1}$, which also leads to  $\kappa_n\ll\kappa_g$ and $\tau\ll\kappa_g$. Eqs. \eqref{eFG} are thus simplified into
\begin{equation} \label{efg_simple}
\begin{aligned}
    &f=D\omega(-4 \kappa_g t(t+\frac{\kappa_n^{\prime}}{\kappa_n}) (1+t^2)+2\kappa (1+t^2)^2-4 (1+3t^2)t^{\prime}),\\
    &g=-4D\omega t(1+t^2) \kappa_n.
\end{aligned}
\end{equation}
for overlaid mode and 
\begin{equation} \label{efg_simple2}
\begin{aligned}
    &f=D\omega(-8 \kappa_g t^2 (1+t^2)+2\kappa (1+t^2)^2-4 \kappa_g(1+3t^2)t^{\prime}),\\
    &g=0,
\end{aligned}
\end{equation}
for symmetric mode.

\begin{figure}[!h]
    \centering
    \includegraphics[width=\linewidth]{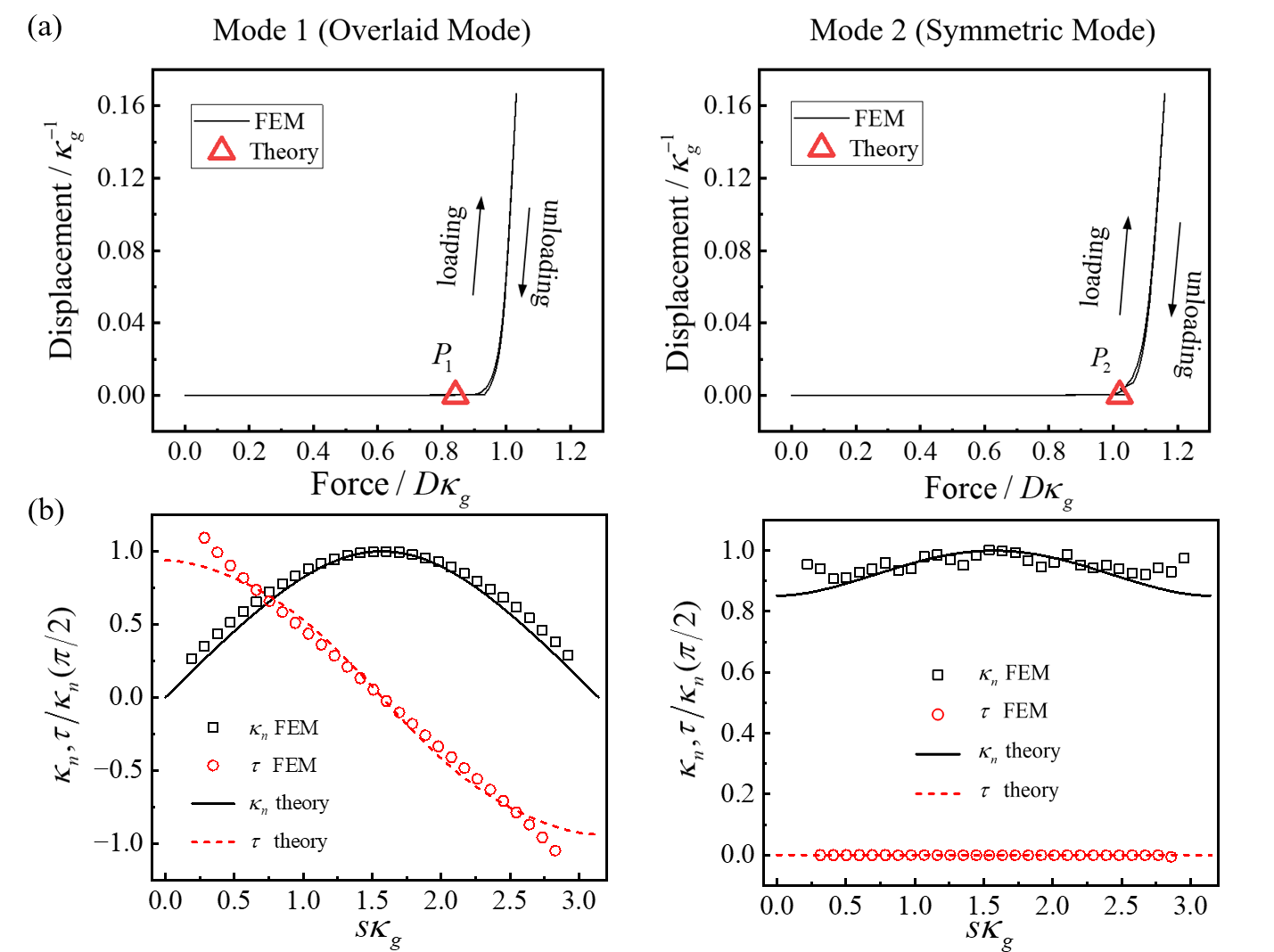}
    \caption{(a) Force-displacement curves for loading and unloading of mode 1 and mode 2. Red triangles represent the predicted critical loads. (b) The distributions of the curvature and  torsion of the crease in mode 1 and mode 2.}
    \label{fig:FEMxy}
\end{figure}
Since $\kappa_{n}/\kappa_g= o(1)$ and $\tau/\kappa_g= o(1)$, from Eq. \eqref{efg_simple} we derive that $f\kappa_g^{-1} =O(1)$ and $g\kappa_g^{-1} =o(1)$. Eqs. \eqref{eEL} are thus simplified into
\begin{equation} \label{EL_simple}
\begin{aligned}
    &\kappa_g f^{\prime}+\frac{f^{\prime\prime\prime}}{\kappa_g}=0\\
    &-\kappa_g g^{\prime}-\frac{g^{\prime\prime\prime}}{\kappa_g}+f\tau^{\prime}+2\tau f^{\prime}=0.
\end{aligned}
\end{equation}

For mode 1 (overlaid mode), the crease has a nonzero torsion. Equations \eqref{eBC}, \eqref{efg_simple} and \eqref{EL_simple} form a boundary value problem which can be solved numerically. The normal curvature $\kappa_n$ and the torsion $\tau$ (i.e., the shape of the crease) are calculated numerically and shown in Fig. \ref{fig:FEMxy}, which are compared with the discrete curvature and the torsion obtained from the FEM results under small deformation ($\Delta y \sim 10^{-2}\kappa_g^{-1}$). We also derive the energy $\delta E$ and displacement $\delta y$ based on the shape of the crease.
%\begin{equation}\label{em1}
%    \begin{aligned}
 %        &\kappa_n=A\kappa_g(\frac{\pi}{2}-(S\kappa_g-\frac{\pi}{2})\sgn(S\kappa_g-\frac{\pi}{2})),\\
  %       &\tau=A\kappa_g\sgn(S\kappa_g-\frac{\pi}{2}).
   % \end{aligned}
%\end{equation}
%where $A$ is an undetermined coefficient with $A=o(1)$. From Eqs. \eqref{em1}, the displacement $\delta y$ and the increased energy $\delta E$ is
%\begin{equation}
 %   \begin{aligned}
  %      &\delta y = A^2\kappa_g^{-1}, \\
   %     &\delta E = A^2\frac{\pi^3}{12} D \omega \kappa_g.
   % \end{aligned}
%\end{equation}
The critical load is thus derived by
\begin{equation}\label{eP1}
    P_1=\frac{\delta E}{\delta y}\approx 0.809 \pi D \omega \kappa_g^2
\end{equation}

For mode 2 (symmetric mode), due to the symmetry, the crease remains planar, i.e., $\tau=0$. From Eqs. \eqref{efg_simple2} and \eqref{eBC}, Eqs. \eqref{EL_simple} are simplified into
\begin{equation} \label{em2_simple}
    \begin{aligned}
        &1-2t^2-3t^4-2\kappa_g^{-1}(1+3t^2)t^{\prime}=\alpha \cos(s\kappa_g)
     \end{aligned}
\end{equation}
where $t=\kappa_n^{\prime}/\kappa_n\kappa_g$ and $\alpha$ is selected to satisfy the boundary condition $\partial_{\kappa^{\prime}}\varepsilon=4D\omega t (1+t^2)=0$, i.e. $t(0)=t(\pi\kappa_g^{-1})=0$. Similarly, we solve the boundary value problem numerically to obtain the normal curvature and torsion (Fig. \ref{fig:FEMxy}). Also, the critical load can be obtained by
\begin{equation}\label{eP2}
    \begin{aligned}
        P_2=\frac{\delta E}{\delta y}\approx 0.974\pi D \omega \kappa_g^2.
    \end{aligned}
\end{equation}
Notably, if $\kappa_n=const$, i.e. the crease remains circular and the flanks remain conical, we have $P_2=\pi D \omega \kappa_g^2$, which is larger than the true $P_2$, meaning that the energy minimization leads to a non-circular crease.

The force-displacement curves of loading and unloading for both modes simulated by FEM are shown in Fig. \ref{fig:FEMxy}, where the predicted critical loads from Eq. \eqref{eP1} and \eqref{eP2} are marked, showing good agreement with the FEM results.
The results reveal that the critical load of the overlaid mode is smaller than the critical load of the symmetric mode. Heuristically, the reason is that the total energy is a functional of the crease shape. The crease in the overlaid mode has a larger configuration space ($\tau \neq 0$), and thus results in a lower energy for the same load. Accordingly, the buckling critical load is smaller.

\section{Bistrip model and large deformation analysis}
\begin{figure}[!h]
    \centering
    \includegraphics[width=\linewidth]{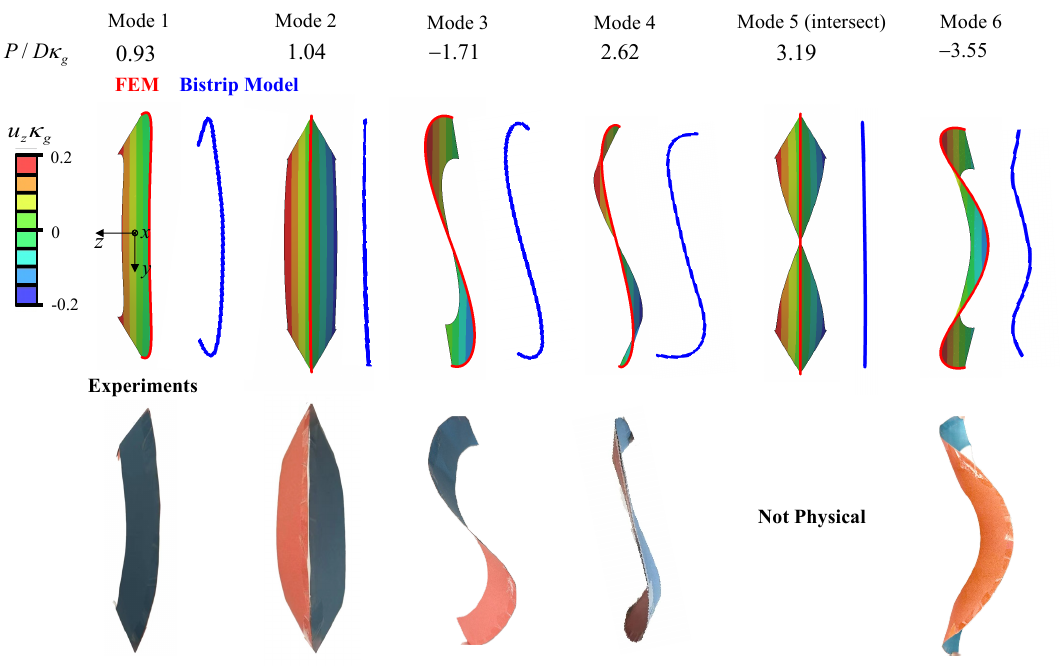}
    \caption{FEM simulations, bistrip models and experiments of deformed curved-fold origami at large deformations. Negative/positive forces denote tension/compression. Mode 5 has intersections of surfaces which is not physically realistic.}
    \label{fig:post}
\end{figure}

In FEM and experiments, we find that more complex modes emerge when the applied force is large (Fig.~\ref{fig:post} modes 3-6).  
Here we propose a preliminary bistrip model based on anisotropic Kirchhoff's rods to analyze these large deformations efficiently. The curved-crease origami is modeled as a single rod by combining two anisotropic rods. In numerical experiments, we realize that the elastic constants, geometries, initial guesses and boundary conditions in the bistrip model will dramatically influence the large deformation modes. Therefore, due to the length limit of this paper, we only present some qualitative results here that capture the main features observed in FEM. A comprehensive study of the model is reserved for future work.

Several recent studies have demonstrated the accuracy of anisotropic Kirchhoff's rods in predicting the nonlinear mechanics of strips with $width/thickness$ up to $ O(10)$ \cite{YU2019657,riccobelli2021rods,moulton2018stable}. 
To develop a bistrip model, we assume the strips are thin and narrow, and each strip is modeled as an anisotropic Kirchhoff's rod under geometric constraints ($\kappa_{g_i} = \kappa \cos\alpha_i$).
The model takes the curved-fold origami as a single rod effectively, characterized by the shape of its crease.
Let the curvature and torsion of the crease be $\kappa$ and $\tau$. Following the derivations in Appendix \ref{sec:app} and \cite{DIAS201457}, we obtain the local equilibrium equations as
\begin{equation}\label{eq:bvp}
    \begin{aligned}
       \kappa'' &= f_1(\kappa, \kappa', m_{12}, m_{22},\bfN),\\ 
\tau' &= f_2(\kappa, \kappa',  m_{12}, m_{22}, \bfN),  \\
m'_{12}&= f_3(\kappa, \kappa',\tau,  m_{12}, m_{22}, \bfN), \\
m'_{22}&= f_4(\kappa, \kappa',\tau,  m_{12}, m_{22}, \bfN),  
    \end{aligned}
\end{equation}
where $m_{i2}$ corresponds to the Lagrangian multiplier of the constraint $\kappa_{g_i} = \kappa \cos\alpha_i$, $f_i$ denotes the explicit function given by Eq.~(\ref{eq:momentequilibrium}),  $\bfN$ is the internal force, and the derivative is taken with respect to the arclength of the crease. Then we use the solve$\_$bvp function in Python to solve the two-point boundary value problems (\ref{eq:bvp}) under the boundary conditions $(\kappa_a = \kappa_b, \kappa'_a = - \kappa'_b, \tau_a = \pm \tau_b, {m_{12}}_a, {m_{22}}_a)$ and the applied force $\bfN=\bfN_0=(N_1,N_2,N_3)$ in Frenet frame of the deformed crease. Here $a$ and $b$ denote the start and end points of the crease respectively. ${m_{12}}_a$ and  ${m_{22}}_a$ are determined by the vanishing global moment in Eq. (\ref{eq:globalmoment})  at the two ends. $\bfN_0$ is in the Frenet frame of the deformed crease.

The results are shown in Fig.~\ref{fig:post}. We choose boundary conditions to capture the main features of the four large deformation modes with elastic constants $EI = GJ = 1$. The applied forces $\bfP$ in FEM are presented, where "$+$" denotes compression and "$-$" denotes tension. The deformed creases from the bistrip model are labeled in blue to compare with the results from FEM. In our bistrip model, Mode 1 and Mode 2 are small deformation results with nonzero and zero torsion ($\bfN_0=(-0.05,0,0)$ and $(-0.08,0,0)$).Modes 3 and 4 have $\tau_a = \tau_b$, exhibiting a similar $S$ shape but responding to different types of forces ($\bfN_0 = (0.1,0.01,0)$ and $(-0.12,-0.01,-0.02)$ respectively). Mode 5 is torsion-free and depicts an unphysical intersection of surfaces ($\bfN_0=(0.08,0,0)$). Mode 6 features changes in the sign of torsion, with the boundary condition $\tau_a = -\tau_b$ and $\bfN_0=(0.23,0,0.02)$.

Although the results show that the bistrip model is capable of capturing the main features of large deformations, some key problems remain open. First, our model does not take the developability condition (zero Gaussian curvature) into account, while a previous strip model does \cite{DIAS201457}. Our model is simplified, but how accurate the model is requires more investigation. Second, the applied force $\bfN_0$ is in the local Frenet frame of the deformed crease, thus it is hard to know the force in global coordinates before finishing the calculation. That is why Euler angles are usually prescribed at two ends for this type of problem \cite{yu2021cutting,YU2019657}, however, the boundary conditions of Euler angles will lack the information of external forces. Third, a continuation analysis (e.g. AUTO) is needed to study the bifurcation behaviors in terms of geometries.

\section{Discussion}
In this work, we derive a unified geometric description of curved-crease origami that bears nontrivial Gaussian curvature at the crease, and then study the mechanical response within this framework. In particular, we investigate the Euler-type buckling of the curved folds, 
relying on theoretical linear stability analysis at small deformations. For large deformation analysis, we develop a numerical bistrip model based on anisotropic Kirchhoff's rods. Our main contributions are summarized as follows.

(1) We derive a unified geometric description of non-Euclidean curved-crease origami in Frenet frame, revealing that four configurations emerge (which will degenerate to two for the symmetric case) for the prescribed reference strips and deformed curved crease, owning to the nonzero Gaussian curvature at the crease.  

(2) We study the mechanics of the curved-crease origami by deriving the local equilibrium equations, with a focus on the Euler-type buckling of the degenerate origami. Both the theoretical prediction and the FEM result indicate that the overlaid configuration possesses a lower buckling threshold. 

(3) We develop a bistrip model based on anisotropic Kirchhoff's rod to study the large deformation behavior. The result of the numerical bistrip model can capture the main features efficiently.

For future research, the bistrip model can be examined more quantitatively, with an emphasis on accuracy and feasibility, to further explore the bifurcation behaviors of such origami.
Developing a dynamic model for this type of origami would be highly valuable, and an inverse design strategy is essential for practical applications.
We believe our results can help understand both the geometry and mechanics of non-Euclidean curved-crease origami, and pave the way for applying curved-fold origami in fields such as robotics, actuators, and space structures.

\appendix
\section{Boundary conditions for the Euler-Lagrange equations} \label{sec:appA}
Since the previous work \cite{WEN2024105829} has not discussed the boundary conditions of the Euler-Lagrange equations Eqs. \eqref{eEL}, here we provide how we derive Eqs. \eqref{eq:bvp}.

According to Eqs. (30), (35), (36) in Wen et al. \cite{WEN2024105829}, the energy variation is given by 
\begin{equation}\label{app1}
\begin{aligned}
\delta E &= \int_{0}^{s_0}(g \kappa \mathbf{b} \cdot \delta \mathbf{r}_c^{\prime}-\frac{g\kappa^{\prime}}{\kappa^2} \mathbf{b} \cdot \delta \mathbf{r}_c^{\prime \prime}+(f-\frac{g\tau}{\kappa}) \mathbf{n} \cdot \delta \mathbf{r}_c^{\prime \prime}+\frac{g}{\kappa}\mathbf{b} \cdot \delta \mathbf{r}_c^{\prime \prime \prime}) \mathrm{d} s\\
&+ \partial_{\kappa}\varepsilon \mathbf{n} \cdot \delta \mathbf{r}_c^{\prime \prime}\bigg |_{0}^{s_0}.
\end{aligned}
\end{equation}
Integrating by parts, Eq. \eqref{app1} is transformed into
\begin{equation}\label{app2}
\begin{aligned}
\delta E &=  \int_{0}^{s_0} (f_{gc} \mathbf{t}+g_{gc} \mathbf{b})\cdot \delta \mathbf{r}_c \mathrm{d} s\\
&+(\partial_{\kappa^{\prime}}\varepsilon\mathbf{n}+\frac{g}{\kappa}\mathbf{b})\cdot \delta \mathbf{r}_c^{\prime\prime}\bigg |_{0}^{s_0}\\
&+(f \mathbf{n}-\frac{g^{\prime}}{\kappa}\mathbf{b})\cdot \delta \mathbf{r}_c^{\prime}\bigg |_{0}^{s_0}\\
&+((-\frac{f}{\kappa} \tau^2+\frac{f^{\prime\prime}}{\kappa}+g\tau+\frac{g^{\prime}}{\kappa^2}\tau^{\prime}+2\frac{\tau}{\kappa} (\frac{g^{\prime}}{\kappa})^{\prime})\mathbf{t}
-(\frac{\tau g^{\prime}}{\kappa}-f^{\prime})\mathbf{n}
-(f\tau+\frac{g^{\prime}\kappa^{\prime}}{\kappa^2}-\frac{g^{\prime\prime}}{\kappa})\mathbf{b})\cdot\delta \mathbf{r}_c\bigg |_{0}^{s_0}.
\end{aligned}
\end{equation}
According to the energy minimum principle, $\delta E=0$ should be satisfied for arbitrary $\delta \mathbf{r}_c$. Since both ends are constrained, $\delta \mathbf{r}_c(0)=\delta \mathbf{r}_c(s_0)=0$. Therefore, the fourth term of Eq. \eqref{app2} automatically equals zero. On the other hand, the boundary values of $\delta \mathbf{r}_c^{\prime}$ and $\delta \mathbf{r}_c^{\prime\prime}$ are arbitrary, leading to that
\begin{equation}\label{eBC2}
\partial_{\kappa^{\prime}}\varepsilon=0,\ \frac{g}{\kappa}=0,\ f=0,\ \frac{g^{\prime}}{\kappa}=0,
\end{equation}
should be satisfied at both ends $s=0,s_0$.
\section{A bistrip model based on anisotropic Kirchhoff's rods}\label{sec:app}
Several recent studies have demonstrated its accuracy in predicting the nonlinear mechanics of strips with $width/thickness$ up to $ O(10)$ \cite{YU2019657,riccobelli2021rods,moulton2018stable}. Throughout this study, we fix $width/thickness$ to the order of magnitude $O(10)$ and assume the geodesic curvature of the crease is frozen. We further assume the two thin strips are inextensible and unshearable, whose force and moment balances can be summarized as  
\begin{equation}\label{eq:n&mbalance}
\begin{aligned}
\bm{n}_1' +\bm{p} &=\bm{0} \,, \bm{m}_1' +\bm{d}_{13} \times \bm{n}_1 +\bm{q}&=\bm{0} \,, \\
\bm{n}_2' -\bm{p} &=\bm{0} \,, \bm{m}_2' +\bm{d}_{23} \times \bm{n}_2 -\bm{q}&=\bm{0} \,, \\
\end{aligned}
\end{equation}
where $\bm{n}_i$ ($=n_{i1}  \bm{d}_{i1} +n_{i2}  \bm{d}_{i2} +n_{i3}  \bm{d}_{i3}$), $\bm{m}_i$ ($=m_{i1}  \bm{d}_{i1} +m_{i2}  \bm{d} _{i2}+m_{i3}  \bm{d}_{i3}$), and $(\bm{ d}_{i 1}, \bm{d}_{i 2}, \bm{d}_{i 3} )$ correspond to the contact forces, contact moments, and the orthonormal right-handed material frame (Darboux frame) attached to the centerline of the $i_{\text{th}}$ strip ($i=1,2$), respectively. $\bm{d}_{i 1}$, $\bm{d}_{i 2}$, and $\bm{d}_{i 3}$ are aligned with the width, surface normal, and local tangent direction. $\bm{p}$ and $\bm{q}$ correspond to the distributed forces and moments that the two strips apply to each other. We have $\bm{d}_{i j} '=\bm{\omega}_i \times \bm{d}_{i j} $ ($i=1,2$ for a bistrip structure; $j=1, 2, 3$). $\bm{\omega}_i= \omega_{i1}  \bm{d}_{i1} + \omega_{i2}  \bm{d}_{i2} +\omega_{i3}  \bm{d}_{i3}$ corresponds to the Darboux vector of the material frame attached to the $i_{\text{th}}$ rod.

Since the forces and moments that the two strips apply to each other are internal forces/moments, we define $\bm{N}=\bm{n}_1 +\bm{n}_2$ and $\bm{M}=\bm{m}_1 +\bm{m}_2$. Adding the equilibrium equations of the two strips together, we have

\begin{equation}\label{eq:N&Mvector}
\begin{aligned}
\bm{N}' =\bm{0} \,, \bm{M}' +\bm{D}_3 \times \bm{N}&=\bm{0} \,, \\
\end{aligned}
\end{equation}

We define the Frenet-Serret frame of the crease as $(\bm{D}_1, \bm{D}_2, \bm{D}_3)$ ($=(\bfn, \bfb, \bft)$ in Section \ref{sec:geometry}), with $\bm{D}_3 = \bm{d}_{i3}$ ($i=1,2$) and $\bm{D}_i' = \bm{\Omega} \times \bm{D}_i$. $\bm{\Omega} = \kappa \bm{D}_2 + \tau \bm{D}_3$ is the Darboux vector, with $\kappa$ and $\tau$ corresponding to the curvature and torsion of the crease. By resolving $\bm{N}$ and $\bm{M}$ onto the crease frame, i.e. , $\bm{N}=N_1 \bm{D}_1 + N_2 \bm{D}_2 +N_3 \bm{D}_3$ and $\bm{M}=M_1 \bm{D}_1 + M_2 \bm{D}_2 +M_3 \bm{D}_3$, and substituting $\bm{D}_i' = \bm{\Omega} \times \bm{D}_i$ into Equation \eqref{eq:N&Mvector}, we have

\begin{equation}\label{eq:N&Mcomponent}
\begin{aligned}
N_1'-N_2 \tau+N_3 \kappa=0 \, , N_2'+N_1 \tau=0 \, , N_3'-N_1 \kappa=0 \, , \\
M_1'-M_2 \tau-N_2+M_3 \kappa =0 \, , M_2'+M_1 \tau+N_1 =0 \, , M_3'-M_1 \kappa =0 \,, \\
\end{aligned}
\end{equation}

The local material frames $(\bm{d}_{i1}, \bm{d}_{i2}, \bm{d}_{i3} )$ can be obtained by rotating the frame $(\bm{D}_1, \bm{D}_2, \bm{D}_3)$ through an angle $\alpha_i =\cos^{-1} (\kappa_{gi} /\kappa )$ ($ \kappa_{gi} \le \kappa $), where $\kappa_{gi} $ corresponds to the frozen geodesic curvature of the $i_{\text{th}}$ rod. We have

\begin{equation}\label{eq:framerela}
\begin{aligned}
\bm{d}_{i1} &= \bm{D}_1 \cos \alpha_i + \bm{D}_2 \sin \alpha_i \,, \\
\bm{d}_{i2} &= \bm{D}_2 \cos \alpha_i - \bm{D}_1 \sin \alpha_i \,, \\
\bm{d}_{i3} &= \bm{D}_3 \,. \\
\end{aligned}
\end{equation}
Differentiating Equation \ref{eq:framerela} with respect to arc length $s$, we have 

\begin{equation}\label{eq:strainrela}
\begin{aligned}
\omega_{i1} = \kappa \sin \alpha_i \,, \omega_{i2} = \kappa \cos \alpha_i =\kappa_{gi} \,, \omega_{i3} = \tau + \alpha_i' \,. \\
\end{aligned}
\end{equation}

The global moment $\bm{M}$ can be obtained by projecting the local moment $\bm{m}_i$ onto the crease frame. The local moment $\bm{m}_i$ can be written as 

\begin{equation}\label{eq:localmoment}
\begin{aligned}
\bm{m}_i &= E_i I_{i1} ( \omega_{i1}-\omega_{i0} ) \bm{d}_{i1} +m_{i2} \bm{d}_{i2} + G_i J_{i} \omega_{i3} \bm{d}_{i3}  \,, \\ 
\end{aligned}
\end{equation}
where we assume linear constitutive laws, and $E_i$ and $G_i$ are the Young's modulus and shear modulus, respectively. $E_i I_{i1} $ and $G_i J_{i1}$ correspond to the bending and torsional rigidity, respectively. $\omega_{i0}$ corresponds to the rest curvature of the strip in the bendable direction. $m_{i2}$ corresponds to the bending moment in the unbendable direction. In other words, it is the Lagrange multiplier associated with the constraint $\omega_{i2}-\kappa_{gi}=0$. The component of the global moment $\bm{M}$ can be written as 

\begin{equation}\label{eq:globalmoment}
\begin{aligned}
M_1 &= ( { \bm{m}_1 +\bm{m}_2 } ) \bm{D}_1 =  \sum _{i=1} ^{2}
[E_i I_{i1} ( \kappa \sin \alpha_i -\omega_{i0} ) \cos \alpha_i - m_{i2} \sin \alpha_i ]  \,, \\ 
M_2&= ( { \bm{m}_1 +\bm{m}_2 } ) \bm{D}_2 =  \sum _{i=1} ^{2}
[E_i I_{i1} ( \kappa \sin \alpha_i -\omega_{i0} ) \sin \alpha_i + m_{i2} \cos \alpha_i ]  \,, \\ 
M_3&= ( { \bm{m}_1 +\bm{m}_2 } ) \bm{D}_3 =  \sum _{i=1} ^{2}
G_i J_{i} (\tau + \alpha_i' )  \,, \\ 
\end{aligned}
\end{equation}

Another constitutive law comes from the mechanics of the crease, which could generate a distributive moment $\bm{q}$ to the two strips. By subtracting the two moment equilibrium equations in \eqref{eq:n&mbalance}, we have 

 \begin{equation}\label{eq:creasemoment}
 \begin{aligned}
(\bm{m}_1 -\bm{m}_2)  ' +\bm{D}_3 \times (\bm{n}_1 -\bm{n}_2) + 2 \bm{q}&=\bm{0}. \\
 \end{aligned}
 \end{equation} 
 Projecting the above equation about $\bm{D_3}$ leads to  
 
  \begin{equation}\label{eq:creasemomentprojection}
 \begin{aligned}
 &(\bm{m}_1 -\bm{m}_2)  ' \bm{D}_3  + 2 q_3 =\bm{0} \,, \\
 \Rightarrow &(\alpha_1 '' - \alpha_2 '' ) - [a ( \kappa \sin \alpha_1 -\omega_{10} ) \cos \alpha_1 - m_{12} \sin \alpha_1 - a ( \kappa \sin \alpha_2 -\omega_{20} ) \cos \alpha_2 + m_{22} \sin \alpha_2 ] \kappa \\&+2 K (\alpha_1 - \alpha_2 -\theta_0)=0 \,,
 \end{aligned}
 \end{equation} 
 where we assume the crease behaves like a linear rotational spring with a spring stiffness $K$ and rest angle $\theta_0$.  
 We further substitute Equation \eqref{eq:globalmoment} into the moment equilibrium in Equation \eqref{eq:N&Mcomponent}, and together with the constraint $\kappa \cos \alpha_i =\kappa_{gi} $ and Equation \eqref{eq:creasemomentprojection},we can solve $\kappa''$, $\tau'$, $m_{12}'$, and $m_{22}'$ as following
 
   \begin{equation}\label{eq:momentequilibrium}
 \begin{aligned}
 \kappa ''=&( \left[  (a ( \kappa \sin \alpha_1 -\omega_{10} ) \cos \alpha_1 - m_{12} \sin \alpha_1 - a ( \kappa \sin \alpha_2 -\omega_{20} ) \cos \alpha_2 + m_{22} \sin \alpha_2)  \kappa \right.
 \\ &-2 K (\alpha_1 - \alpha_2 -\theta_0) \left. \right] \kappa \sin \alpha_1 \sin \alpha_2 
 +2 \kappa' \sin \alpha_1 \sin \alpha_2 (\alpha_1' - \alpha_2' ) + \kappa \cos \alpha_1 \sin \alpha_2 (\alpha_1')^2 \\ &- \kappa \cos \alpha_2 \sin \alpha_1 (\alpha_2')^2 + \kappa_{g1}'' \sin \alpha_2 - \kappa_{g2}'' \sin \alpha_1 ) / \sin( \alpha_2-\alpha_1 ) \,, \\ \\
 \tau'=&[ M_1 \kappa^2  \sin \alpha_1 \sin \alpha_2 - \kappa'' \sin( \alpha_2+\alpha_1 ) + 2 \kappa' \sin \alpha_1 \sin \alpha_2 (\alpha_1' + \alpha_2' ) + \kappa \cos \alpha_1 \sin \alpha_2 (\alpha_1')^2 \\
 &+ \kappa \cos \alpha_2 \sin \alpha_1 (\alpha_2')^2 + \kappa_{g1}'' \sin \alpha_2 + \kappa_{g2}'' \sin \alpha_1 ) ] / (2 \kappa \sin \alpha_1 \sin \alpha_2) \,, \\ \\
 m_{12}'=& m_{12}  \alpha_1'  \cot (\alpha_2-\alpha_1) +\frac{m_{22} \alpha_2'}{\sin  (\alpha_2-\alpha_1)}  - a ( \kappa' \sin \alpha_1 +\kappa \alpha_1' \cos \alpha_1  ) \cot (\alpha_2-\alpha_1) - a ( \kappa \sin \alpha_1 - \omega_{10}  ) \alpha_1'  \\
 &  - [a ( \kappa' \sin \alpha_2 + \kappa \alpha_2' \cos \alpha_2  ) + (M_3 \kappa -M_2 \tau -N_2) \cos \alpha_2 + (M_1 \tau +N_1 ) \sin \alpha_2] / \sin (\alpha_2 - \alpha_1 ) \,, \\ \\
 m_{22}'=&\frac{m_{12}  \alpha_1'} {\sin (\alpha_1-\alpha_2)} +m_{22} \alpha_2'  \cot  (\alpha_1-\alpha_2)  - \frac{a ( \kappa' \sin \alpha_1 +\kappa \alpha_1' \cos \alpha_1  )} {\sin (\alpha_1-\alpha_2)} - a ( \kappa \sin \alpha_2 - \omega_{20}  ) \alpha_2'  \\
 &  - a ( \kappa' \sin \alpha_2 + \kappa \alpha_2' \cos \alpha_2  ) \cot(\alpha_1 -\alpha_2) -  \frac{ (M_3 \kappa -M_2 \tau -N_2) \cos \alpha_1 + (M_1 \tau +N_1 ) \sin \alpha_1}  {\sin (\alpha_1 - \alpha_2 )} \,, \\
 \end{aligned}
 \end{equation} 
which can be solved numerically as a boundary value problem.

\section*{Acknowledgments} 
The authors thank the
National Natural Science Foundation of China (grant no. 12472061) for support of this work.
\bibliographystyle{unsrt}
\bibliography{curved_fold}

\end{document}